\documentstyle[epsfig]{mn2e}{}
\begin{document}

\def\mpc{h^{-1} {\rm{Mpc}}}
\def\up{h^{-3} {\rm{Mpc^3}}}
\def\uk{h {\rm{Mpc^{-1}}}}
\def\lsim{\mathrel{\hbox{\rlap{\hbox{\lower4pt\hbox{$\sim$}}}\hbox{$<$}}}}
\def\gsim{\mathrel{\hbox{\rlap{\hbox{\lower4pt\hbox{$\sim$}}}\hbox{$>$}}}}
\def\kms {\rm{km~s^{-1}}}
\def\apj {ApJ}
\def\aj {AJ}
\def\mnras {MNRAS}
\def\aap {A\&A}

\title[Large-scale modulation of star formation in void walls]{Large-scale modulation of star formation in void walls}
\author[Ceccarelli, L., Padilla, N., Lambas, D.G.]{
\parbox[t]{\textwidth}{Ceccarelli, L.$^{1}$, Padilla, N.$^{2}$, Lambas, D.G.$^{1}$
}
\vspace*{6pt}\\
$^1$ IATE, Observatorio Astron\'omico de C\'ordoba, Argentina. \\
$^2$ Departamento de Astronom\'\i a y Astrof\'\i sica, Pontificia
     Universidad Cat\'olica de Chile, Santiago, Chile.\\
}

\date{\today}

\maketitle

\begin{abstract}
We perform a statistical study of the characteristics of galaxies in voids
and void walls in the SDSS and 2dFGRS catalogues.  
We investigate dependencies of the distribution of 
galaxy spectral types and colours as a function of the relative position to the
void centres for different luminosity and local density ranges.  We find a trend
towards bluer, star forming galaxies in void walls beyond the local density dependence.
These results indicate that luminosity and local density do not entirely determine
the distribution of galaxy properties such as colours and spectral types, and
point towards a large scale modulation of star formation.  We argue that this effect
is due to the lower accretion and merger history of galaxies arriving at void walls from
the emptier inner void regions.
\end{abstract}


\section{Introduction} 
The study of the galaxy population in large voids is crucial to understand 
the processes involved in the formation and evolution of galaxies.
Voids have been found to occupy the greatest ammounts of space in the universe, adding up to
$\sim 40\%$ of the total volume of galaxy surveys (Patiri et al., 2005). 
This result has been repeatedly found using several catalogues at a variety of 
wavelengths (Hoyle \& Vogeley, 2002), such as 
the Center for Astrophysics Survey (CfA, Vogeley et. al, 1991; 1994), 
the Southern Sky Redshift Survey (SSRS, Gazta\~naga \& Yokohama, 1993), 
the Point Source Catalogue Redshift Survey (PSCz, Hoyle \& Vogeley, 2002), 
the Infrared Astronomical Satellite (IRAS, El-Ad et. al, 1997), 
the Las Campanas Redshift Survey (LCRS, Muller et. al, 2000),
the 2degree Field Galaxy Redshift Survey (2dFGRS, Hoyle \& Vogeley, 2004; 
Croton et. al, 2004; Patiri et. al, 2005) and 
the Sloan Digital Sky Survey (SDSS, Hoyle et. al, 2005; Rojas et. al, 2005).

Even though voids occupy such a large fraction of the volume of the Universe,
the very low number density of galaxies inside them (lower than $10\%$ the average
in the Universe) makes it difficult to obtain samples of void galaxies suitable
for statistical measures.  This is another reason why voids need to be studied in 
the largest possible surveys covering wide solid angles to considerable depths.

Several studies have been focused on the properties of galaxies in underdense 
regions.  The luminosity function of galaxies in voids in the Sloan Digital Sky Survey
has been measured by Rojas et al. (2005), who also study their photometric properties
finding that in general the population of galaxies in voids is characterised by
a fainter characteristic luminosity $L^*$ although the relative importance
of faint galaxies is similar to that found in the field (i.e. the faint-end
slope of the luminosity function in voids is similar to that of the field).
Spectroscopic properties of void galaxies have also been 
studied in detail (Hoyle, Vogeley \& Rojas, 2005); these results indicate 
that galaxies inside voids 
have higher star formation rates than galaxies in denser regions 
and are still forming starts at the same rate than in the past. 

Another widely used application of void statistics is that of probing the bias in
the galaxy distribution using large redshift surveys (Mathis \& White 2002; 
Arbabi-Bidgoli et al. 2002; Benson et al. 2003; Goldberg \& Vogeley 2004).
Statistical studies of voids have also been found to provide invaluable information on 
higher order clustering (White 1979; Fry 1986) which can be used to probe models
of galaxy clustering (Croton et al., 2004) and void properties such as sizes, shapes 
and frequency of occurrence, and how these properties vary with galaxy type.
Void statistics in general, have been shown to provide important clues on the galaxy formation
processes and can be used to place constraints on cosmological models (Peebles 
2001).
   \begin{figure}
   \epsfxsize=6.5cm
   \vspace*{-.2cm}
   \hspace*{-.1cm}
   \centerline{\epsffile{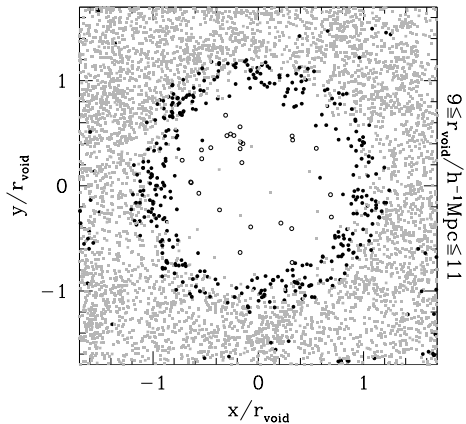}
   }
   \vspace*{-.6cm} 
   \caption{
   Stacked distribution of galaxies around SDSS voids
   of radius $9\leq r_{\rm void}/$h$^{-1}$Mpc$\leq11$; wall galaxies are
   plotted as black, filled circles, inner void galaxies are represented by open circles,
   and galaxies at distances $>1.15r_{\rm void}$ are
   shown as grey crosses.
    }
   \label{fig:voidgx}
   \end{figure}

   \begin{figure}
   \epsfxsize=7.5cm
   \vspace*{-.1cm} \centerline{\epsffile{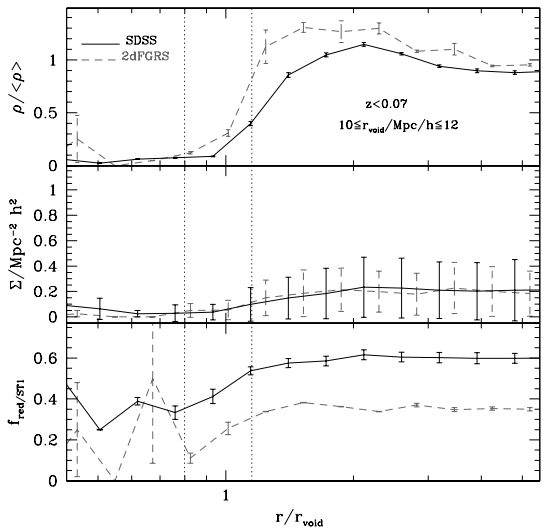}
   }
   \vspace*{-.2cm}
   \caption{
   Upper panel: void density profiles traced by all galaxies in the SDSS (solid lines) and the 2dFGRS (dashed lines), for
   galaxies with $M_r<-19.2$ (SDSS) and $M_{bJ}<-18$ (2dFGRS).  Void radii are in the range 
   $10\leq r_{\rm void}/$h$^{-1}$Mpc$ \leq 12$.
   Middle panel: average local galaxy density as a function of normalised distance
   to the void centres.
   Lower panel: fraction of galaxies with $\eta<-1.3$ in the 2dFGRS and
   of red galaxies in the SDSS, selected using $g-r<2.2$.
    }
   \label{fig:profiles}
   \end{figure}
Voids are also found in numerical simulations, either identified using the 
distribution of dark-matter particles, dark-matter haloes, or semi-analytic galaxies.
Large scale cosmological simulations and mock galaxy catalogues 
have been used to study void properties, including their 
evolution, distribution of sizes and density profiles  
(Benson et al. 2003; Goldberg \& Vogeley 2004; Sheth \& van de 
Wiegaert 2004; Patiri et al. 2004; Shandarin et al. 2004; 
Colberg et al 2005; Padilla, Ceccarelli \& Lambas 2005 and references therein).  
Other studies have focused on the properties of halos populating the inner 
volumes of voids (Antonuccio-Degelou et al. 2002; Gottoberg et al. 2003; 
Goldberg \& Vogeley 2004; Patiri et al. 2004; Goldberg et al. 2005), finding
that inside voids, the mass function becomes consistent with that of a lower density Universe.

Recent observational and theoretical results suggest that 
large underdense regions generate coherent outflows of mass and galaxies 
moving toward the void edges (Ceccarelli et al., 2006, Padilla et al., 2005), thus it is
possible that galaxies arriving at the void edges or
walls could have experienced a different evolutionary history than their field counterparts.  For instance
due to the void material accumulating around them,
or to the fact that void galaxies most likely spent their lives inside voids.  
Motivated by these facts, 
in this letter we will perform a statistical study of
wall and normal galaxies using the 2dFGRS and SDSS,
analising spectroscopic and photometric properties 
of galaxies, taking into account the well known dependence on
luminosity/stellar mass and local density
(Balogh et al., 2004, Baldry et al., 2006, Dekel \& Birnboim, 2006, 
Kannappan 2004, Lagos et al., 2008).

\section{Data samples}
The samples analised in this work are taken from the Sloan Digital Sky Survey, Data
Release 6 (SDSS DR6, Adelman-McCarthy et al., 2007) and the 2-degree Field Galaxy Redshift
Survey (Colless et al., 2003).  The use of both catalogues allows us to probe a larger volume
of space to improve the robustness of our results by using two independent
surveys covering different regions of the sky.

We select the $\sim 585,000$ galaxies from the $r < 17.77$ magnitude-limited main
spectroscopic galaxy sample from the
SDSS DR6 (Adelman-McCarthy et al., 2007), which contains 
CCD imaging data in five photometric bands ($ugriz$,
Fukugita et al., 1996).
We will use to define this high quality photometry to define 
galaxy colours; in particular we will use the $u-r$ since
it has proven to be more sensitive to variations in galaxy properties and environment
(e.g. Baldry et al., 2006).  
    \begin{figure*}
    \epsfxsize=10cm
    \vspace*{-.6cm} 
    \centerline{\epsffile{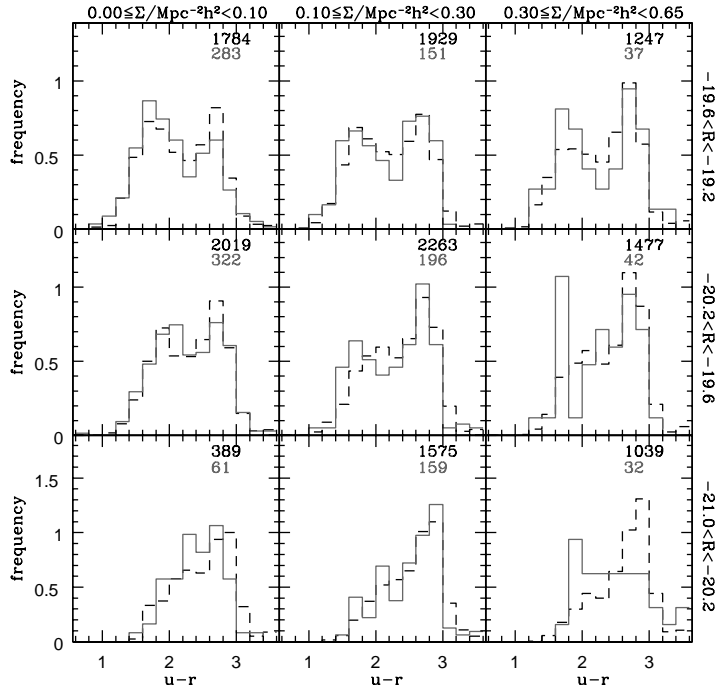}
    }
    \vspace*{-.45cm} 
    \caption{
Distributions of $u-r$ colour index for galaxies in different luminosity and
local density bins as indicated in the key.  Solid lines correspond to the field 
SDSS sample (number of galaxies in each subpanel in black) and
dashed lines to SDSS galaxies in void walls, $0.8<r/r_{\rm void}<1.15$ (number of
galaxies in grey).
     }
    \label{fig:ur}
    \end{figure*}
The 2dFGRS contains spectroscopic redshifts for approximately $230,000$ galaxies.
The source catalogue for the 2dFGRS is a revised and extended version of 
the APM galaxy catalogue from which a set of target galaxies, characterised by 
extinction-corrected magnitudes brighter than b$_J$=19.45, was selected for the
construction of the 2dFGRS.  In all our analysis we use a magnitude limit b$_J$=18.9
to avoid the variation of completeness with angular position.  In order to 
quantify the properties of void galaxies we use the spectral 
clasification introduced by Madgwick et. al (2002), included in the catalogue.
Every galaxy is asigned a spectral parameter $\eta$, which represents the average
line strenght in its spectrum. The spectral parameter ($\eta$) shows a strong correlation with
the equivalent width of the H$\alpha$ emision line and can therefore be interpreted 
as an indicator of the ammount of star formation
in the galaxy, which allows to define a spectral clasification.
The specific star formation rate in galaxies becomes higher as $\eta$ increases.
Given that 2dFGRS plate photometry is of less quality than SDSS CCD data we use
$\eta$ rather than $b_J-R$ colours to characterise the star formation
activity in 2dFGRS galaxies.

\section{Voids in galaxy catalogues}
We apply the void finding algorithm described in Ceccarelli et al. (2006, hereafter C06),
which tests the density inside spheres centred on a very large number of random
centres for a wide range of sphere radii.
For each centre the largest sphere that satisfies a low density criterion 
becomes a void candidate (note that only a small fraction of the initial, random 
candidate centres will satisfy the low density condition).
As a final step we remove small spheres contained in larger ones.

We have adopted $z=0.08$ 
in the SDSS and 2dFGRS as the limiting redshift of our
samples, so that the catalogues are complete for galaxies brighter than 
$M_r=-19.2$ and $M_{bJ}=-18$ 
in SDSS and 2dFGRS, respectively.  This choice of maximum redshift is a compromise between
well-resolved voids which require faint galaxies, and a sufficiently large volume in order
to have enough void statistics.
The adopted absolute magnitude limits imply that the galaxy number density is high enough
to lower the effects of shot noise in the identification of small voids.
    \begin{table*}
    \begin{center}
    \begin{tabular}{cccccc}
    \hline
    \hline
    \noalign{\vglue 0.2em}
    sample  & $z_{lim}$ & Max. Lum. & No. of gals & No. of voids & No. of wall gals. \\
    \noalign{\vglue 0.2em}
    \hline
    \noalign{\vglue 0.2em}
     $SDSS$ &  $0.08$ & $M_r<-19.2$ & $66849$ & $136$& $2674$\\
     $2dFGRS$ &  $0.08$ & $M_{bJ}<-18.0$ & $26654$ &$39$ & $1432$\\
    \noalign{\vglue 0.2em}
    \hline
    \hline
    \end{tabular}\label{table:glxsamples}
    \vskip -0.2cm
    \caption{Characteristics of galaxy and void samples.}
    \end{center}
    \end{table*}
We apply the void finding algorithm to the samples of galaxies described in table 1. 
Our resulting samples of voids are restricted to radii within the
range $5$ to $15$h$^{-1}$Mpc, which comprises the best resolved systems
suitable for our study (the resulting number of voids are shown in table 1).  
Fig. 1 shows the stacked distribution of galaxies around voids 
in the SDSS-DR6; the galaxy positions are in units of their closest
void radius, and we restrict the $z-$coordinate to lie 
within $0.5r_{\rm void}$ of the void centre to allow
a better visualization. 
In addition to the simple check allowed by this figure, 
the reliability of the identified voids in all these samples 
has been carefully tested in C06.

\section{Galaxies in void walls}

In order to study the effects of the evolutionary history of galaxies arriving at the void edges
we define the void walls as the spherical shells delimited by distances to void
centres of $0.8$ and $1.15r/r_{\rm void}$ (galaxies in walls are shown as black solid circles
in Fig. 1; the upper limit marks the beggining of the decrement in the fractions or red galaxies).

\subsection{Galaxy densities: global vs. local}
Several papers have analysed the relative fraction of galaxy populations of
different characteristic colour and morphology as a function of environment.  The fact
that local galaxy density may be crucial in determining the properties of galaxies has
been explored by many authors since the pioneering work of Dressler (1982; see
for instance, Balogh et al., 2004, 
Baldry et al., 2006, Rojas et al., 2004, Patiri et al., 2005,
Hoyle et al., 2005).  Given this well documented dependence, we perform
a local density dependent analysis in order to analyse the properties of galaxies in void walls
with respect to galaxies of similar local densities not residing in void walls.  
This analysis should able to determine the relative weights of global and 
local effects on galaxy properties.

We define a local density parameter, $\Sigma$, as the projected number density of galaxies brighter than
$M_r\leq-20.2$ 
in the SDSS and $M_{b{\rm J}}<-19.1$ in the 2dFGRS, within 
projected distances of $d<2.5$h$^{-1}$Mpc, and radial velocity differences of
$\Delta V=1000$km/s.  We use this fixed radius rather than the usual $\Sigma_5$ calculated
using the fifth nearest neighbour, in order to avoid non-local effects at low density environments.
As can be seen in Figure 
\ref{fig:profiles} (upper panel),
the average, non-local density in void walls is of a few tenths of the average
density of galaxies in the full catalogue, in both the SDSS or the 2dFGRS.  The middle
panel shows the mean local density of galaxies (errorbars show the dispersion around the mean) 
as a function of void-centric distance;
note that galaxies in void walls span a wide range in local densities, from nearly isolated
to group/cluster galaxies, with similar values in the 2dFGRS and SDSS catalogues.  
The lower panel shows the relative fraction of 2dFGRS type 1 galaxies,
corresponding
to red, passively star forming objects ($\eta<-1.3$), 
and the fraction of red galaxies ($u-r>2.2$)
in the SDSS-DR6;
The differences in the fractions shown in the 2dFGRS and SDSS are expected since
the 2dFGRS is based on a bluer colour selection than the SDSS. It can be seen that at the edge
of voids ($0.8<r/r_{\rm void}<1.15$), there is a systematic drop in these fractions, which
we will explore into more detail in the following sections, since this is expected to some
degree given the well known relation between spectral morphology and colours with local
density.  

\subsection{Bimodality of the galaxy distribution}
    \begin{figure}
    \epsfxsize=7.5cm
    \vspace*{-.7cm}
    \centerline{\epsffile{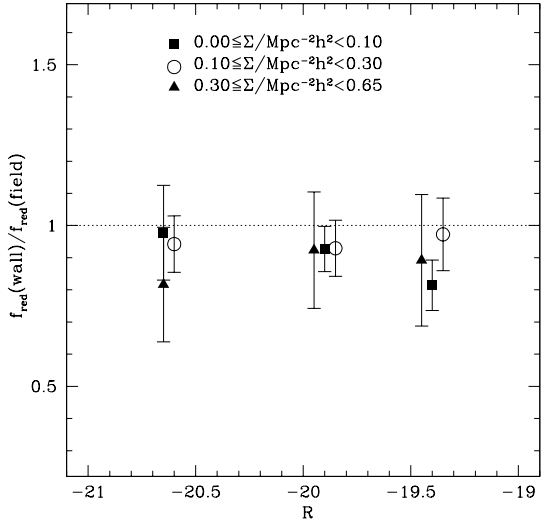}
    }
   \vspace*{-.2cm}
    \caption{
Red galaxy fractions for wall galaxies in low and high local density environments relative
to galaxies outside voids.  Results are shown for different galaxy luminosities (see the figure key).
    }
    \label{fig:redf}
    \end{figure}
    \begin{figure}
    \epsfxsize=9cm
    \vspace*{-0.7cm}
    \hspace*{0.0cm} \centerline{\epsffile{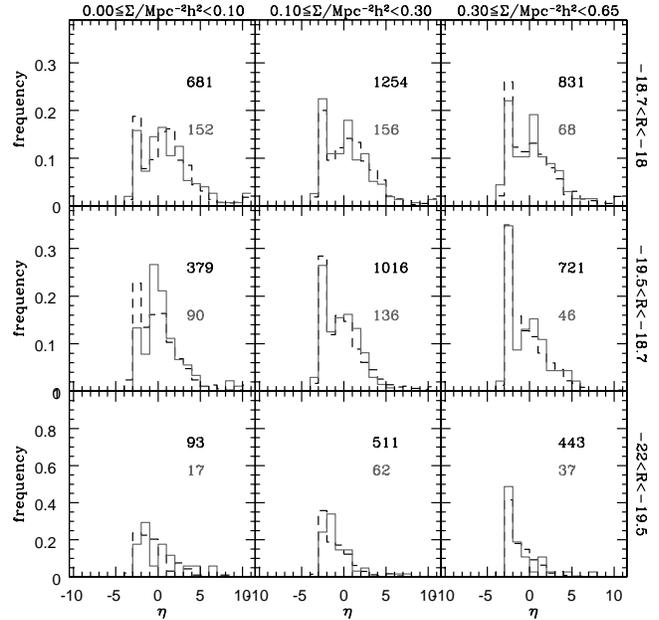}
    }
   \vspace*{-.5cm}
    \caption{
Distributions of spectral type parameter $\eta$ for 2dFGRS galaxies for different luminosity
and local density ranges.  Line types are as in Fig. \ref{fig:ur}
     }
    \label{fig:eta}
    \end{figure}
    \begin{figure}
    \epsfxsize=7.5cm
    \vspace*{-.7cm}
    \centerline{\epsffile{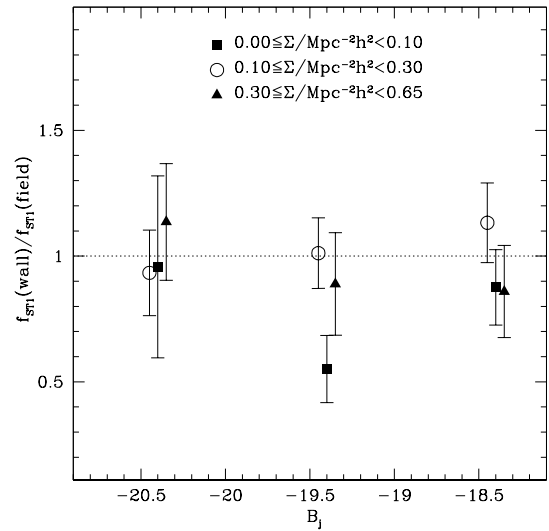}
    }
   \vspace*{-.2cm}
    \caption{
Fractions of galaxies with $\eta<1.3$ for wall in low and high local density environments relative
to galaxies outside voids.  Results are shown for different galaxy luminosities (see the figure key).
    }
    \label{fig:t1f}
    \end{figure}
Given that the bimodal behaviour of galaxies is a strong function of luminosity
and local density, we have studied the distribution of SDSS-DR6 $u-r$ colours as a function
of these two variables for galaxies in void walls and outside voids separately.  Thus,
any differences in these distributions can only be related to the astrophysical effects
associated to the special star formation history of galaxies which today reside within void walls.

The results are shown in figure \ref{fig:ur}, where it can be appreciated that the full sample
of galaxies shows the well documented effect that brighter galaxies of equal local density
occupy preferentially the red peak of the colour distribution.  However, we find a systematic
trend of galaxies in void walls to be bluer at any given luminosity and local density values.

This can be seen more clearly in the fraction of red galaxies as a function of luminosity for
the different local density ranges explored.  The results are shown in Fig. \ref{fig:redf}.  
We notice that the red fractions in void wall galaxies are systematically lower than galaxies outside
voids by up to $-0.18\pm0.18$ for the higher luminosity galaxies in high density reginos, and by up to $-0.19\pm0.08$ 
for faint galaxies in low local densities.  Although with a lower
statistical significance, this is also the case for galaxies with higher local densities.
The global ratios adding together galaxies in the different luminosity bins provide higher
significance results, where low densities (corresponding to the solid squares in the figure) 
show a ratio between wall and field samples of
$f_{\rm red}({\rm wall})/f_{\rm red}({\rm field})=0.89\pm0.05$, and high densities (corresponding to the solid
triangles)
$f_{\rm red}({\rm wall})/f_{\rm red}({\rm field})=0.88\pm0.11$, a $2-$ and $1-\sigma$ detections, respectively.

We have also explored this effect using 2dFGRS data using the $\eta$ parameter.
Similar differences between void wall galaxies and the field
can be appreciated in Figs. \ref{fig:eta} and \ref{fig:t1f},
where we show the distributions of $\eta$ for different luminosities and local density bins, and the
resulting ratios between late-type and total galaxy populations for wall and field catalogues.
Although the significance is lower than in the SDSS, an
excess of star-forming galaxies is seen for void wall galaxies particularly 
in low luminosity, low local density
environments consistent with the results shown in Fig. \ref{fig:ur}.  This reinforces the
statistical significance of our findings by including the largest combined 
spectroscopic sample available to date.
\vskip -2cm
\section{CONCLUSIONS}
In this work we have focussed on an analysis of properties of galaxies residing in void edges,
aimed at exploring possible differences induced by the 
different interaction history of galaxies arriving at void walls driven by the wall expansion.
Given the strong dependence of
galaxy colour index and spectral type on luminosity and local density,
we have considered different ranges in these two parameters to analyse the properties
of the population of void wall galaxies compared to galaxies outside voids.

Our analysis indicates that galaxies residing in void walls are systematically bluer and
more actively star-forming at a given luminosity and local galaxy density range.
These results suggest that besides the influence of local environment,
galaxies are also subject to a large scale dependent star-formation activity; in the
case studied here, by the lower interaction history of galaxies escaping void interiors.  
This is an effect
taking place over scales of the order of void radii, which in this study corresponds to $5-15$h$^{-1}$Mpc.

One important aspect that needs to be studied with larger samples of voids is whether this effect
is also present at the more internal void regions, which may also show different properties
to what is simply expected from the local density effects on galaxy properties.  

This finding of a large scale modulation of star formation can be used to test galaxy formation
scenarios, and adds an extra parameter in the relation between galaxy properties and 
environment.

\section*{Acknowledgments}
This work has been partially supported by Consejo de Investigaciones 
Cient\'{\i}ficas y T\'ecnicas de la Rep\'ublica Argentina (CONICET), the
Secretar\'{\i}a de Ciencia y T\'ecnica de la Universidad Nacional de C\'ordoba
(SeCyT), Fundaci\'on Antorchas, Argentina and Agencia C\'ordoba Ciencia, Centro
de Astrof\'\i sica FONDAP. 
NP was supported by Proyecto Fondecyt Regular no. 1071006.
LC was partially supported by the Latin-american and European Network 
for Astrophysics and Cosmology (LENAC) Project. DG beneffited from travel funds
from Proyecto Fondecyt de Incentivo a la Cooperaci\'on Internacional no. 7070045.

\end{document}